\documentclass[10pt,preprint]{aastex}
\usepackage{psboxit}

\PScommands

\newcommand{\latin}[1]{{#1}}

\newcommand{\eg}{\latin{e.g.}}

\def\simless{\mathbin{\lower 3pt\hbox
	{$\,\rlap{\raise 5pt\hbox{$\char'074$}}\mathchar"7218\,$}}} 
\def\simgreat{\mathbin{\lower 3pt\hbox
	{$\,\rlap{\raise 5pt\hbox{$\char'076$}}\mathchar"7218\,$}}} 

\newcommand{\band}[2]{\ensuremath{^{#1}\!{#2}}}

\newcommand{\uband}{\ensuremath{\band{0.1}{u}}}
\newcommand{\gband}{\ensuremath{\band{0.1}{g}}}
\newcommand{\umg}{\ensuremath{\band{0.1}{(u-g)}}}
\newcommand{\absmg}{\ensuremath{M_{\band{0.1}{g}} -5\log_{10} h}}

\newcounter{thefigs}
\newcommand{\fignum}{\arabic{thefigs}}

\newcounter{thetabs}

\newcounter{address}

\shortauthors{Blanton {\it et al.} (2006)}
\shorttitle{ Quiet years for blue galaxies}

\begin{document}

\title{ Galaxies in SDSS and DEEP2: a quiet life on the blue
sequence?}

\author{
Michael R. Blanton\altaffilmark{\ref{NYU}}}  

\setcounter{address}{1}
\altaffiltext{\theaddress}{
\stepcounter{address}
Center for Cosmology and Particle Physics, Department of Physics, New
York University, 4 Washington Place, New
York, NY 10003
\label{NYU}}

\begin{abstract}
In the six billion years between redshifts $z=1$ and $z=0.1$, galaxies
change due to the aging of their stellar populations, the formation of
new stars, and mergers with other galaxies. Here I explore the
relative importance of these various effects, finding that while
mergers are likely to be important for the red galaxy sequence they
are unlikely to affect more than 10\% of the blue galaxy sequence.  I
compare the galaxy population at redshift $z \sim 0.1$ from the Sloan
Digital Sky Survey (SDSS) to the sample at $z \sim 1$ from the Deep
Extragalactic Evolutionary Probe 2 (DEEP2).  Galaxies are bluer at
$z\sim 1$: the blue sequence by about 0.3 mag and the red sequence by
about 0.1 mag, in \umg. I evaluate the change in color and in the
luminosity functions of the two sequences using some simplistic
stellar population synthesis models. These models indicate that the
luminous end of the red sequence fades less than passive evolution
allows by about 0.2 mag. Due to a lack of luminous blue progenitors,
``dry'' mergers betweeen red galaxies then must create the luminous
red population at $z\sim 0.1$ if stellar population models are
correct. The blue sequence colors and luminosity function are
consistent with a reduction in the star-formation rate since $z\sim 1$
by a factor of about three, with no change in the number density to
within 10\%. These results restrict the number of blue galaxies that
can fall onto the red sequence by any process, and in particular
suggest that if mergers are catastrophic events they must be rare for
blue galaxies.
\end{abstract}


\section{ Motivation}
\label{motivation}

How galaxies form is naturally an important and interesting topic for
astronomers, living as we do in the disk of a relatively large and
typical galaxy. A well-developed theory of galaxy formation exists, in
which galaxies form at the centers of the potential wells of the dark
matter halos that are a natural prediction of the current standard
Cold Dark Matter (CDM) cosmology. Gasdynamic simulations
(e.g. \citealt{kerev05a,springel05b,nagamine05a}) and simpler
``semi-analytic'' models (e.g. \citealt{benson03a,nagashima05a})
attempt to make predictions of this theory.  The class of predictions
I address here involve the change of galaxies (and consequently of the
galaxy population) over time. Of course, given the finite age of the
Universe, galaxies must change over time, but the predictions are more
specific than that and qualitatively point mostly in the same
direction: that star formation was higher in galaxies in the past and
that their luminosities were larger (\citealt{springel03a,
nagamine01a}).

High redshift galaxy surveys of the past ten or twenty years or so
have revealed such changes --- indeed, the observations predate the
theoretical predictions. The first such observations were in clusters,
and indicated that the fraction of blue galaxies has declined markedly
in the last 3--4 Gyrs (\citealt{butcher84a}). Observations of galaxies
in the field have indicated that this process occurs in the field as
well (\citealt{lilly95b, cowie96a, cohen02a, im02a, gabasch04a,
bell04a, faber05a, willmer05a}). Taken as a whole, these observations
have indicated that between redshifts $z=1$ and $z=0$ galaxies became
dimmer and less star-forming.

Here, I present a detailed comparison of the change in the luminosity
and color distribution of galaxies between redshifts around $z\sim 1$
and $z\sim 0.1$.  At redshift $z\sim 0.1$ I use the Sloan Digital Sky
Survey (SDSS; \citealt{york00a}) to measure the color and absolute
magnitude distribution of galaxies. As \citet{blanton03d}
demonstrated, the structure of this distribution consists of the
well-known red sequence of galaxies (\citealt{baum59, faber73a,
visvanathan77, terlevich01a}) and a similar but much broader blue
sequence of galaxies. In this paper, I compare this distribution to
the corresponding distribution in the Deep Evolutionary Extragalactic
Probe 2 (DEEP2; \citealt{davis03a, faber03a}).

In Section \ref{data}, I describe the two data sets. In Section
\ref{convert} I describe the procedure for the conversion of the SDSS
sample into the DEEP2 sample. In Section \ref{results} I compare the
SDSS and DEEP2 samples. In Section \ref{model} I present a simple
model of the results. In Section \ref{discussion}, I discuss the
implications of the models and the data. In Section \ref{conclusions},
I present conclusions.

Where necessary, I have assumed cosmological parameters $\Omega_0 =
0.3$, $\Omega_\Lambda = 0.7$, and $H_0 = 100$ $h$ km s$^{-1}$
Mpc$^{-1}$ (with $h=1$). All magnitudes are (unless otherwise noted)
AB-relative. Throughout, I will refer to a piece of software, {\tt
kcorrect}\footnote{\tt http://cosmo.nyu.edu/blanton/kcorrect}, for
converting between various bandpass systems (Blanton et al. in
preparation).  Except where noted, I will always be referring to the
version of the software labeled {\tt v4\_1\_4}. Finally, note that
most of my comparisons will be in the \band{0.1}{g} band, which is the
SDSS $g$ band blueshifted by a factor 1.1, and has virtues extolled
below.

\section{ Data}
\label{data}

\subsection{SDSS}

The SDSS is taking $ugriz$ CCD imaging of $10^4~\mathrm{deg^2}$ of the
Northern Galactic sky, and, from that imaging, selecting $10^6$
targets for spectroscopy, most of them galaxies with
$r<17.77~\mathrm{mag}$ \citep[\eg,][]{gunn98a,york00a,abazajian03a}.
Automated software performs all of the data processing: astrometry
\citep{pier03a}; source identification, deblending and photometry
\citep{lupton01a}; photometricity determination \citep{hogg01a};
calibration \citep{fukugita96a,smith02a}; spectroscopic target
selection \citep{eisenstein01a,strauss02a,richards02a}; spectroscopic
fiber placement \citep{blanton03a}; and spectroscopic data reduction.
Descriptions of these pipelines also exist in \citet{stoughton02a}.
An automated pipeline called {\tt idlspec2d} measures the redshifts
and classifies the reduced spectra (Schlegel et al., in preparation).

The spectroscopy has small incompletenesses coming primarily from (1)
galaxies missed because of mechanical spectrograph constraints
\citep[6~percent;][]{blanton03a}, which leads to a slight
under-representation of high-density regions, and (2) spectra in which
the redshift is either incorrect or impossible to determine
($<1$~percent).  In this context, I note that the mechanical
constraints are due to the fact that fibers cannot be placed more
closely than 55$''$; when two or more galaxies have a separation
smaller than this distance, one member is chosen independent of its
magnitude or surface brightness. Thus, this incompleteness does not
bias the sample with respect to luminosity. In addition, there are
some galaxies ($\sim 1$~percent) blotted out by bright Galactic stars,
but this incompleteness should be uncorrelated with galaxy properties.

\subsection{NYU-VAGC}
\label{vagc}

For the purposes of computing large-scale structure and galaxy
property statistics, we have assembled a subsample of SDSS galaxies
known as the NYU Value Added Galaxy Catalog (NYU-VAGC;
\citealt{blanton05a}).  Here I use the version of that catalog
corresponding to the SDSS Data Release 4 (DR4;
\citealt{adelman06a}). The reader can obtain this catalog at our web
site\footnote{{\tt http://sdss.physics.nyu.edu/vagc}}.  The main
advantage of this catalog is that it accurately and almost completely
describes the window function of the SDSS, including the flux limit
and completeness as a function of position. Doing so allows
statistical studies of galaxies. In this case, I can estimate the
maximum volume in which SDSS could have observed each galaxy
$V_{\mathrm{max}}$, accounting for the flux limits and completeness of
the survey.

The sample I use here consists of galaxies with Galactic extinction
corrected (\citealt{schlegel98a}), Petrosian magnitudes $14.5 < m_r <
17.6$, and redshifts $0.05 < z < 0.15$.  For calculating
$V_\mathrm{max}$, the evolution correction applied is:
\begin{equation}
E(z) = Q_0 (1 + Q_1 (z - z_0) ) (z-z_0)
\end{equation}
with $z_0 =0.1$, $Q_0= 2.8$ and $Q_1 = -1.8$. For the selection and
$V_\mathrm{max}$ determination I use $K$-corrections from {\tt
kcorrect v3\_4} (\citealt{blanton03b}); note that this is an earlier
version than I use for the rest of the calculations in this paper, but
for these purposes the differences are tiny (percent-level).  In other
respects I select this sample according to the Main sample criteria
that \citet{strauss02a} describe.

When necessary below, we will $K$-correct the magnitudes to \uband\
and \gband\ bands, the SDSS $u$ and $g$ bands shifted to match their
rest frame coverage at $z=0.1$, the typical redshift in our
sample. This choice is wise because these bandpasses also are close in
wavelength to the DEEP2 bandpasses in the rest frame.  For example,
the effective wavelength of \uband\ and \gband\ are 3223 \AA\ and 4245
\AA\ respectively, while for $R$ and $I$ the effective rest frame
wavelengths at redshift $z=1$ are 3297 \AA\ and 4059 \AA\
respectively. Note that \gband\ is also fortuitously close to the
rest frame $B$ band (4344 \AA) of \citet{bessell90a}.

In order to compare our results here to other related results, one can
use the approximate relationships:
\begin{eqnarray}
\band{0.1}{u} &=& u + 0.33 +0.32[(u-g) -1.26] \cr
\band{0.1}{g} &=& g + 0.32 + 0.25[(u-g) -1.26] \cr
\band{0.1}{g} &=& B + 0.07 + 0.06[(B-V) -0.59]
\end{eqnarray}
In particular, note the small color term between \band{0.1}{g} and
$B$.  (All of these magnitudes are AB relative).

\subsection{DEEP2} 

As the redshift $z\sim 1$ sample, I use DEEP2 (\citealt{davis03a,
faber03a}). The DEEP2 team took Canada-France-Hawaii Telescope images
in $B$, $R$, and $I$. \citet{coil04a} describe the construction of a
catalog based on these data. DEEP2 used a set of color criteria
(discussed more fully below) to select galaxies likely to have
redshifts $z>0.7$. Using the DEIMOS spectrograph on Keck 2, they
obtained spectra for $>10,000$ of these galaxies.

DEEP2 has released an initial set of data (DR1).\footnote{\tt
http://deep.berkeley.edu/DR1} These data include the observations in
their 2002 observing season and have 5,191 high quality redshifts
(quality $\ge$ 3), 2,976 of them in the redshift range $0.8 < z <1.2$.

To obtain rest-frame quantities from this data, I use the routine {\tt
deep\_kcorrect} from {\tt kcorrect} to convert to SDSS $\uband$ and
$\gband$ bands, which are close in the rest frame to the $R$ and $I$
DEEP2 observations.  The routine {\tt deep\_kcorrect} fits a simple
spectral model to the $B$, $R$, and $I$ observations, and calculates
for the model the ratio between the rest-frame $\uband$ or $\gband$
band flux and the observed frame $R$ or $I$ band flux.  Then it
multiplies that ratio to the observed $R$ or $I$ flux to obtain the
estimated rest frame $\uband$ or $\gband$ flux. When I use similar
code to convert the observed $R$ band flux to the rest-frame $B$ band
flux, I obtain results very similar to those in Figure A15 of
\citet{willmer05a}, demonstrating that I am using $K$-corrections
similar to those used by the DEEP2 team. 

\subsection{DEEP2 $V_{\mathrm{max}}$ values}
\label{vmax}

For each DEEP2 galaxy in the redshift range $0.8 < z < 1.2$, I also
calculate a maximum volume over which it could be detected:
\begin{equation}
V_{\mathrm{max}} = \frac{1}{3} \int d\Omega \int_{0.8}^{1.2} dz
\frac{d[D_c(z)^3]}{dz} g(z, z_\mathrm{act}, B, R, I)
\end{equation}
where the angular integral is over the full DR1 area, $z_\mathrm{act}$
is the actual redshift in question, and the function $g(z,
z_\mathrm{act}, B, R, I)$ expresses the probability of selecting a
galaxy at the given redshift $z$ given its SED (which I estimate from
$z_{\mathrm{act}}$ and its $B$, $R$, and $I$ magnitudes).

To calculate the integral over angle, I would have liked to have
simply taken geometrical descriptions of the masks and calculated the
area, but the DR1 release did not include descriptions I could
interpret. So I instead took a simpler and approximate approach.  I
counted objects from the full photometric DEEP2 catalog within 45$''$
of any object in an observed DR1 mask, finding a total $N_m$ matched
objects. I then counted the objects in the full DEEP2 photometric
catalog, finding a total $N_p$ over an area $\Omega_p$.  The DEEP2
spectroscopic area is then approximately $\Omega = N_m \Omega_p / N_p
= 1.94\times 10^{-4}$ square radians, or 0.63 sq deg (unweighted by
completemness).

To calculate the integral over redshift, I use a Monte Carlo approach.
For each object, I randomly choose 12,000 values of redshift $z$
between $0.8$ and $1.2$, distributed according to volume. For each
redshift, I calculate what the magnitudes of the object would be using
the {\tt deep\_kcorrect} routinue of {\tt kcorrect}. I then apply the
magnitude and color cuts of DEEP2:
\begin{eqnarray}
B-R &<& 2.35 (R-I) - 0.25\mathrm{\quad, or} \cr
R-I &>& 1.15\mathrm{\quad, or} \cr
B-R &<& 0.50\mathrm{.}
\end{eqnarray}
in addition to $R < 24.1$. 

Furthermore, I apply the estimated incompleteness of DEEP2.
\citet{willmer05a} estimate the magnitude of incompleteness due to
surface brightness effects ($5$--$7$\%), star-galaxy separation errors
($\sim 10\%$), and the unobserved $B$-band dropouts ($4$--$8$\%). I
ignore these effects here. However, I need to account for two
important effects that \citet{willmer05a} do discuss. First, because
of an early change in the selection criteria, the number of targets
selected to be targeted is a function of $I$ magnitude (J. Newman,
private communication). Second, redshift success is a function of
color. Thus, I evaluate as a function of $R$ and $R-I$ the fraction of
possible galaxies that were observed ($f_t(R, R-I)$) and the fraction
of observed galaxies with successful redshifts ($f_g(R, R-I)$). I
obtain results similar to those in Figure 5 of
\citealt{willmer05a}. In fact, according to \citet{willmer05a}, most
of the blue galaxies without successful redshifts in the DEEP2 survey
are at $z>1.4$, so I will simply set $f_g(R, R-I) = 1$ for $R-I <
0.9$. I select random redshifts from the set of 12,000 with
probabilities depending on the predicted observed colors equal to the
product $f_t(R,R-I) f_g(R, R-I)$.

Then the estimate for $V_\mathrm{max}$ for each galaxy is the volume
between redshifts $z=0.8$ and $z=1.2$, times the fraction of the 12,000
selected redshifts which pass the above selection.

\section{ SDSS as it would appear in DEEP2}
\label{convert}

The SDSS catalog contains measurements in the $u$, $g$, $r$, $i$, and
$z$ bands for galaxies at redshifts $z\sim 0.1$. Meanwhile, DEEP2
contains measurements in $B$, $R$, and $I$ for galaxies at redshifts
$z \sim 1$. Furthermore, the SDSS has a simple flux limit in the
$r$-band (there is a surface brightness selection as well, but as
\citealt{blanton04b} show it is unimportant for the range of
luminosities I consider here). Meanwhile, DEEP2 is selected in the $R$
band, which is comparable in rest-frame wavelength to the $u$ band for
the SDSS galaxies. Any comparison between the two surveys must account
for these differences.  I do so here as follows.

First, one must correct the SDSS sample for the flux selection, in
order to be dealing with an effectively volume-limited sample. I do so
by assigning a probability of selection to each galaxy that is
proportional to $1/V_\mathrm{max}$. Then I randomly select a large set
of galaxies (with replacement) from the original sample.

Second, I assign each galaxy a redshift in the range $0.5 < z < 1.5$,
distributed uniformly in comoving volume. Given its redshift I
determine for each selected galaxy what its $B$, $R$, $I$ magnitudes
would be using the routine {\tt sdss2deep} contained in {\tt kcorrect}
(Blanton et al. in preparation). I use the SDSS ``model'' fluxes and
uncertainties to fit a spectral energy distribution (SED) to each
galaxy.  It calculates the flux ratio between each DEEP2 band and the
nearest SDSS band in wavelength (the SDSS bands for the rest-frame at
the actual redshift of the galaxy and the DEEP2 band for the
rest-frame at the assigned redshift). It then multiplies that ratio to
the flux in the nearest SDSS band to obtain an estimate of the DEEP2
flux. The DEEP2 and SDSS filter curves I assume are given in the {\tt
kcorrect} product.

I have tested this method by taking the converted galaxies, applying
to them the DEEP2 $K$-corrections I describe above, and comparing the
resulting \uband\ and \gband\ absolute magnitudes to what one would
find by $K$-correcting directly from the SDSS data. In the mean, these
two methods agree to about 0.03 mag, with about 0.05 scatter. This
test also provides a consistency check on our $K$-corrections of the
DEEP2 galaxies.

Third, I apply the color, magnitude, and incompleteness selection to
the sample, as described in Section \ref{vmax}.

\begin{figure}
\figurenum{\fignum}
\plotone{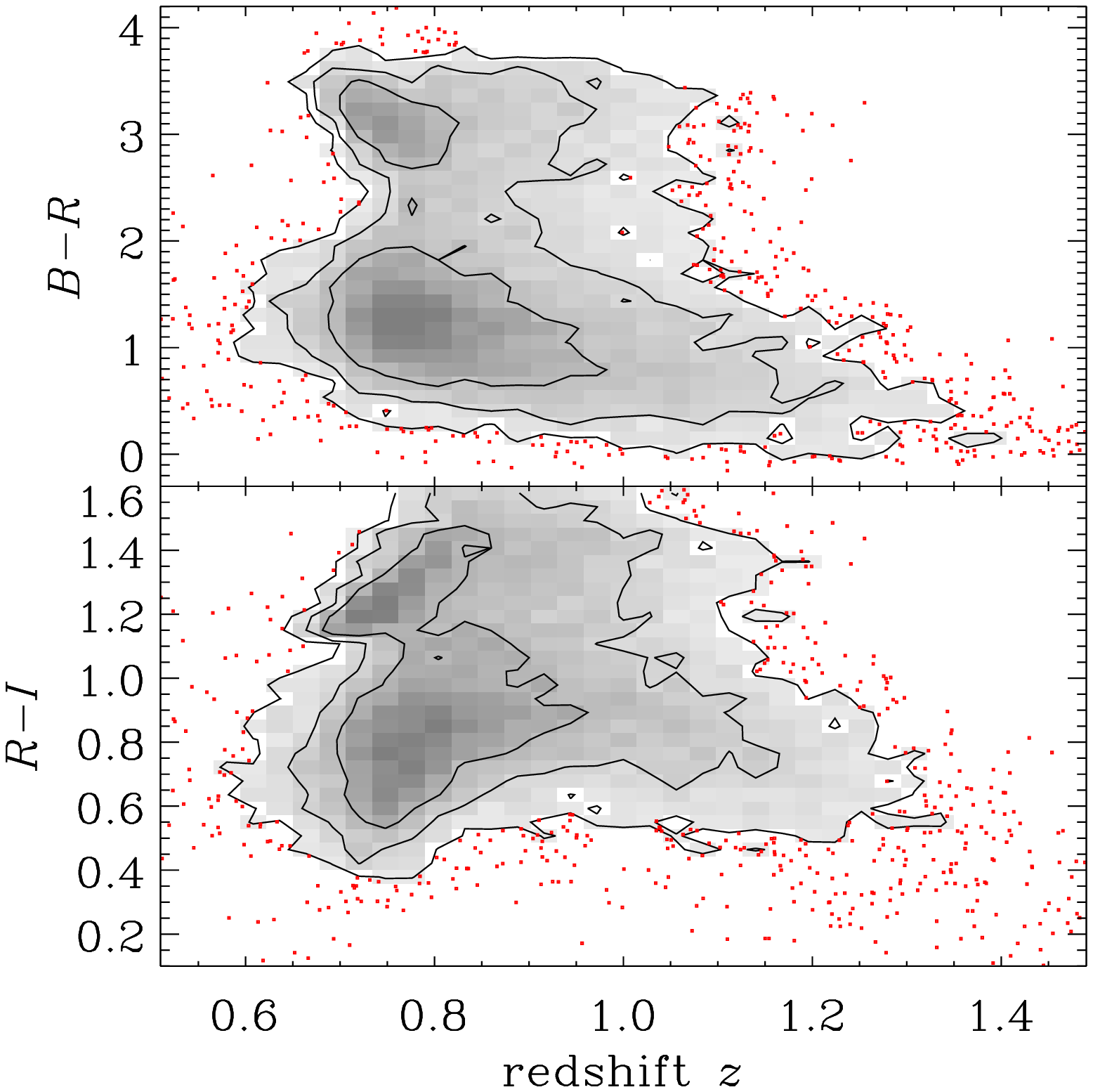}
\caption{\label{noev} Predictions of DEEP2 colors from the local
SDSS sample.  The SDSS galaxies are redistributed in the range $0.5 <
z < 1.5$ uniformly in volume and $B$, $R$, and $I$ magnitudes are
synthesized as described in the text. Greyscale indicates the density
of galaxies in each bin of redshift and color. Contours enclose 52\%,
84\% and 97\% of the galaxies, respectively. We show outlying galaxies
as individual points.}
\end{figure}

Figure \ref{noev} shows the $B-R$ and $R-I$ distributions for the
resulting sample. One can see that, indeed, the distribution cuts off
sharply below redshift $z=0.7$, just as the DEEP2 team intended. In
addition, the red sequence of galaxies declines at the highest
redshifts, purely due to the selection effects.

It is worth noting that the $B$ band at redshift $z=1$ is close to the
near ultra-violet (NUV) band of the Galaxy Evolution Explorer (GALEX),
much bluer than there are observations for most of the SDSS
galaxies. However, Blanton et al. (in preparation) show that the fits
to the SDSS bands correctly predict the GALEX NUV bands to within 0.5
mag scatter and 0.2 mag in the mean. So the clear separation into two
populations in $B-R$ is real.
 
\section{ Change in the galaxy population}
\label{results}

\subsection{ Colors as a function of redshift}

\begin{figure}
\figurenum{\fignum}
\plotone{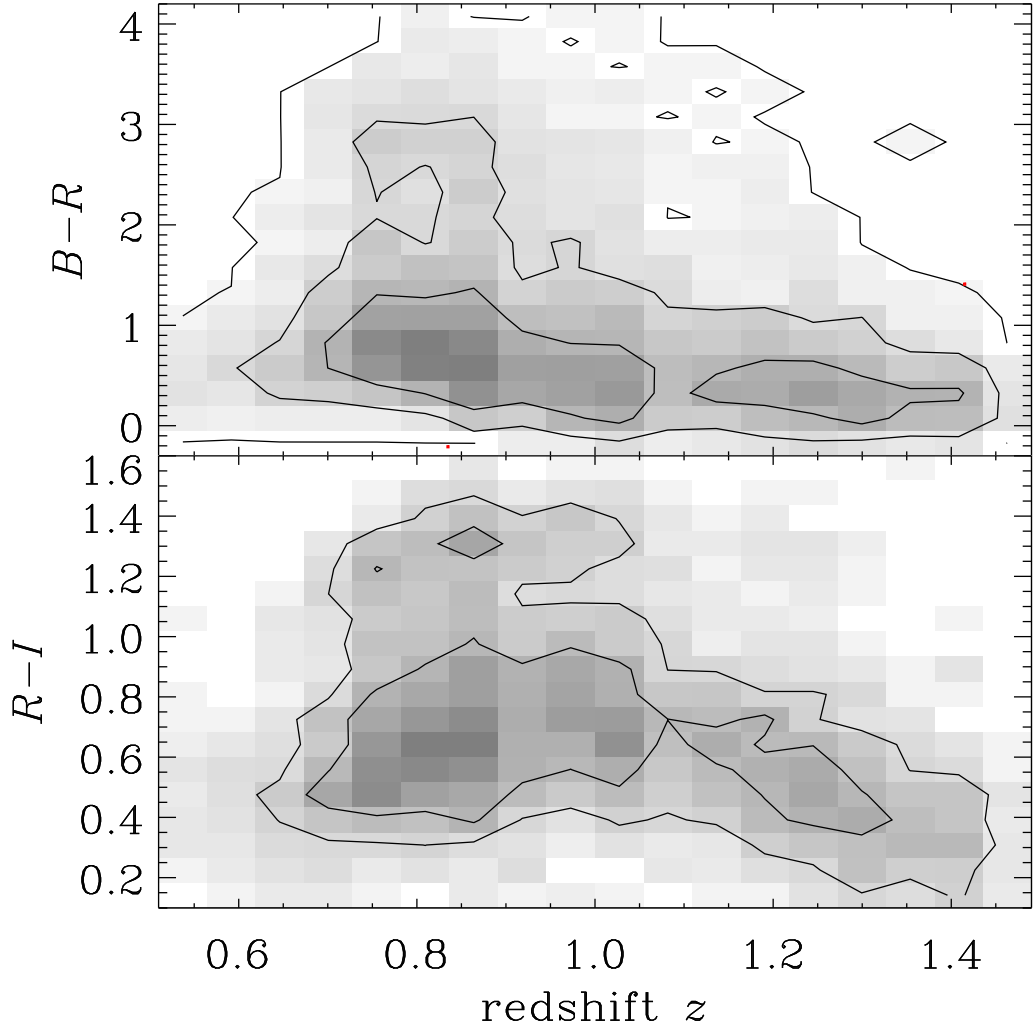}
\caption{\label{deep2} Same as Figure \ref{noev}, for actual DEEP2 
colors as a function of redshift. Obviously the galaxy population is
far bluer than the prediction from low redshift --- meaning galaxies
at high redshift are far bluer than those today. As we show in Figures
\ref{cmag} and \ref{colorhist}, the blue and
red sequences are bluer at redshift $z=1$ than at $z=0.1$, and in
addition that the red sequence is much less well populated at higher
redshift.}
\end{figure}

Figure \ref{deep2} shows the same quantities as Figure \ref{noev}, but
for the actual DEEP2 galaxies.  Clearly there is a huge difference in
the two populations. The DEEP2 sample has a much weaker red sequence
relative to the local population. In addition, its blue sequence is
bluer.

\subsection{ Comparison of the color-magnitude diagram}

To quantify these differences, I take DEEP2 galaxies in the redshift
range $0.8 < z < 1.2$ and correct them to the rest frame \uband\ and
\gband\ bands. For uniformity, I do exactly the same thing to galaxies
in the SDSS prediction (rather than use the SDSS estimates of the
magnitudes directly). Again, for this task I use a routine ({\tt
deep\_kcorrect}) from {\tt kcorrect}.

\begin{figure}
\figurenum{\fignum}
\plotone{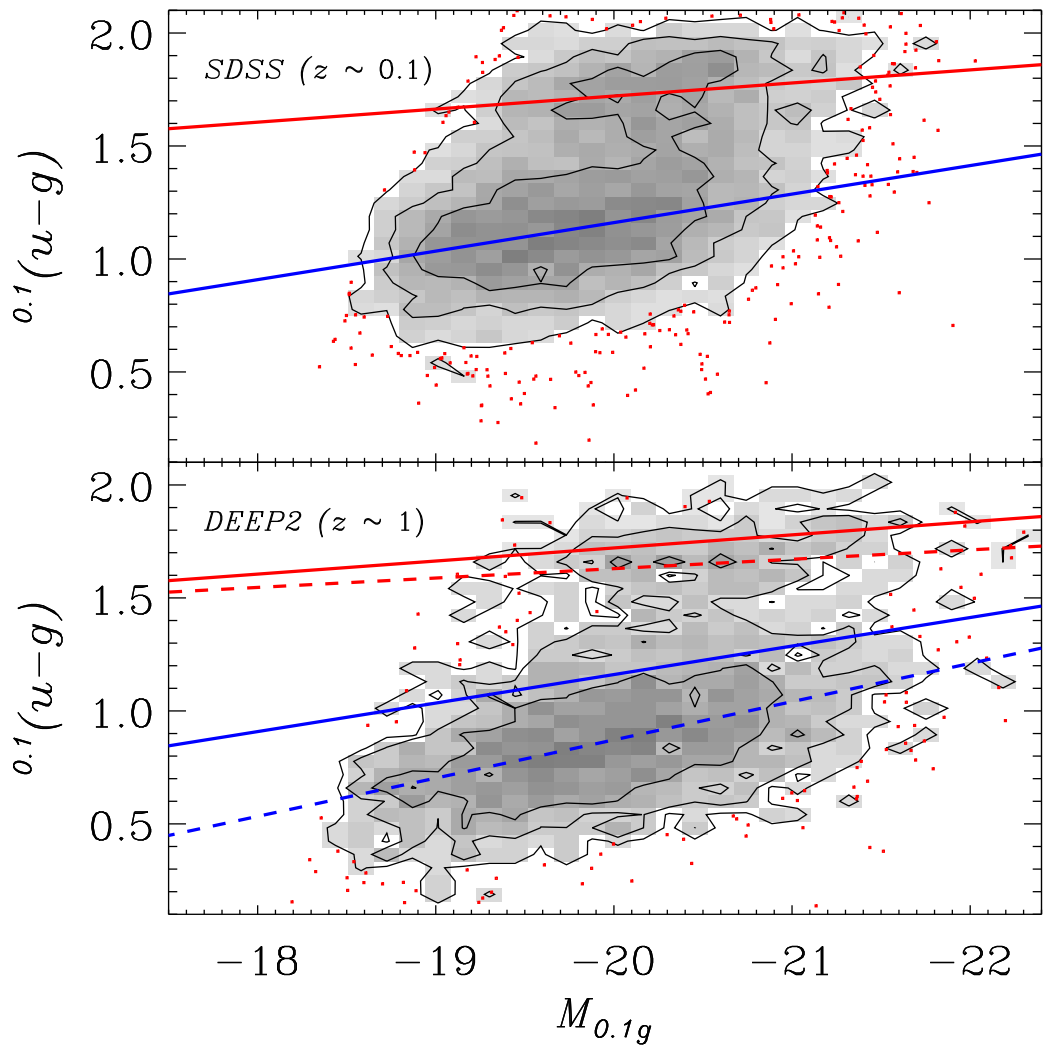}
\caption{\label{cmag} The SDSS-predicted (top panel) and observed
(bottom panel) distribution of color and absolute magnitude.  The
prediction assumes no change in the galaxy population between
redshifts $z=0.1$ and $z=1$. The upper solid line in each panel is the
locus of the red sequence in the SDSS prediction. The lower solid line
is the locus of the blue sequence in the SDSS prediction.  The dashed
lines in the bottom panel are the corresponding loci in the DEEP2
data. The parameters of these fits are all in Table
\ref{cmag_table}. The red sequence is far less well populated relative
to the blue sequence at redshift $z=1$ than it is at redshift
$z=0.1$. In addition, the red and blue sequences are both bluer at
high redshift than at low redshift. }
\end{figure}

Figure \ref{cmag} shows the result: the top panel shows the prediction
from SDSS in the case that there is no change in the galaxy
population; the bottom panel shows the actual DEEP2 observations. The
upper solid line in each plot is the locus of the red sequence in the
SDSS. The lower solid line is the locus of the blue sequence in the
SDSS. The dashed lines are the corresponding fits to the DEEP2
sample. Table \ref{cmag_table} gives the parameters of the fits
to the data.

The red sequence does not shift {\it very} much between SDSS and DEEP2
--- it is about 0.1 mag bluer at high redshift. On the other hand the
blue sequence shifts considerably.  For example, the dashed line is
the locus of the blue sequence in DEEP2, which is 0.3 mag bluer in
\umg\ than the SDSS locus.

\begin{figure}
\figurenum{\fignum}
\plotone{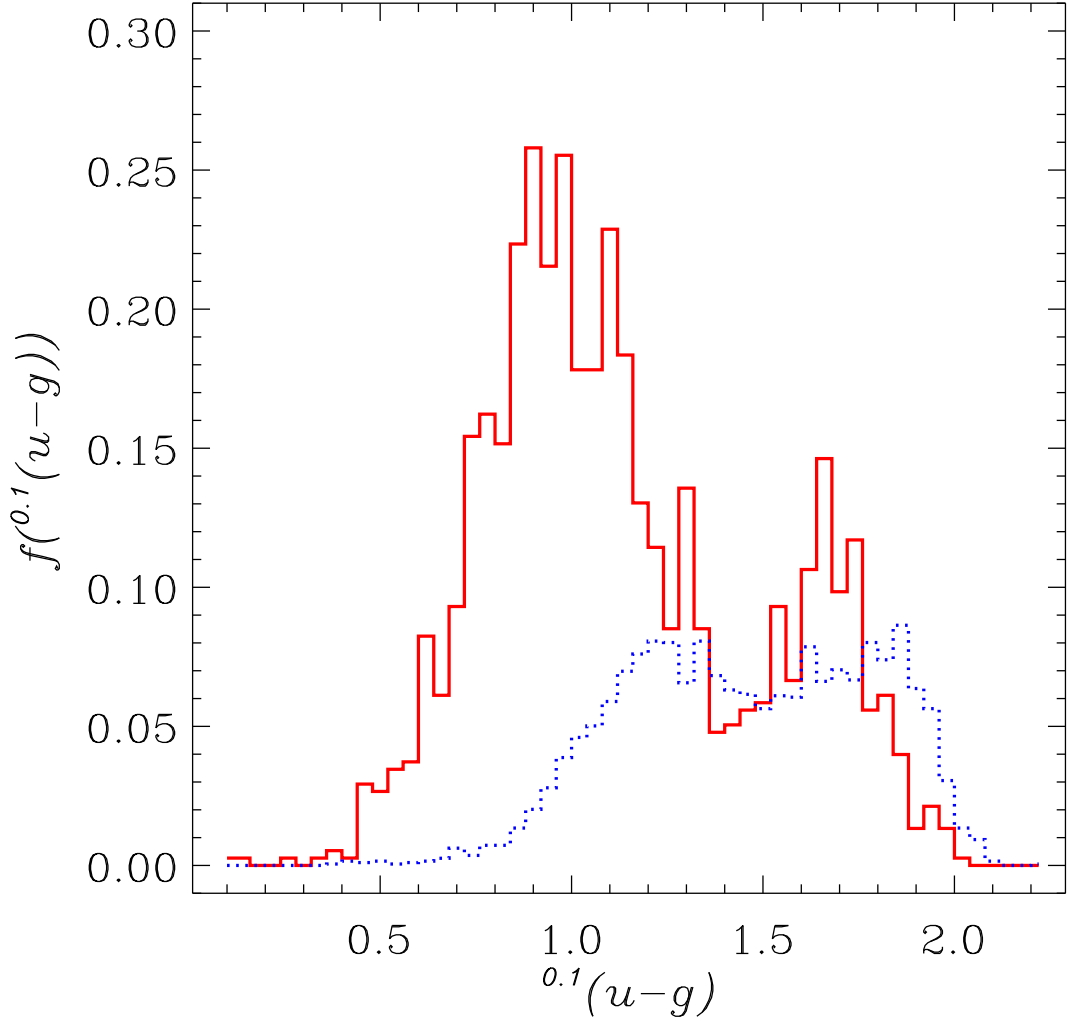}
\caption{\label{colorhist} For luminous galaxies ($\absmg
< -20$), the distribution of DEEP2 colors (solid histogram) compared
to the SDSS prediction assuming no change in the population (dotted
histogram). The curves are scaled such that the integrals are equal
for $\umg > 1.4$. As Figure \ref{cmag} shows, the blue population in
DEEP2 is much bluer than its low redshift analog. In addition, the red
sequence is slightly bluer (about 0.1 mag) and much less populated. In
the DEEP2 sample, galaxies with $\umg>1.4$ comprise 24\% of the sample,
while in the SDSS prediction, galaxies with $\umg>1.5$ comprise 47\% of
the sample. }
\end{figure}

For luminous galaxies ($\absmg < -20$), for which the selection
effects are less important, Figure \ref{colorhist} shows the
distribution of color in both the actual (solid histogram) and the
SDSS-predicted (dotted histogram) samples. Here one clearly sees the
shift in the red and blue sequences, as well as the overwhelming
dominance of the blue sequence at high redshift. The fraction of red
galaxies at redshift $z=1$ is about 0.24, whereas the corresponding
fraction in the redshift $z=0.1$ sample is 0.47. This change is the
effect \citet{butcher84a} discovered, but in the field.

In what follows, I will show that it is reasonable for both the red
and blue sequences to passively evolve. If we approximately account
for that evolution by using a cut $\absmg < -20$ in the SDSS sample,
we find the red fraction is about 0.33, indicating that much of the
shift in the color distribution is due to observing (at fixed absolute
magnitude) lower stellar mass objects at higher redshift.

\subsection{ Comparison of luminosity functions}

While Figures \ref{cmag} and \ref{colorhist} show the raw
distributions of DEEP2 galaxies, I can also attempt to correct the
galaxies for the flux and color limits by weighting each galaxy by
$1/V_\mathrm{max}$ as calculated in Sections \ref{vagc} and
\ref{vmax}. Weighting by this value allows each galaxy to contribute
its number density, and the sum of $1/V_\mathrm{max}$ for some set of
galaxies should thus be the number density of galaxies in that set. I
divide the galaxies into red and blue galaxies for three different
samples:
\begin{enumerate}
\item the original SDSS sample; 
\item the converted SDSS sample, as described in Section
	\ref{convert};  and
\item the DEEP2 sample.
\end{enumerate}
I then calculate the luminosity functions of each sample: the number
density per unit magnitude and volume as a function of absolute
magnitude.

\begin{figure}
\figurenum{\fignum}
\plotone{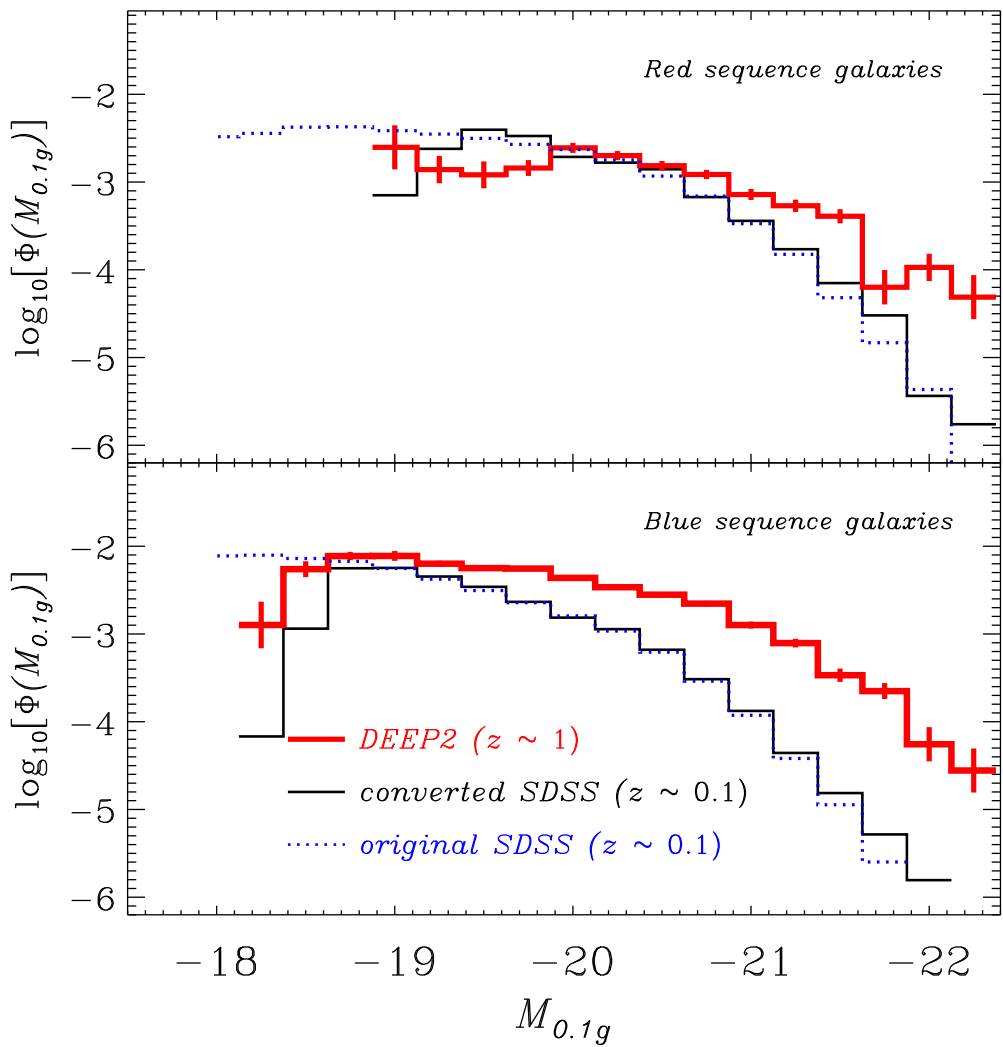}
\caption{\label{lfs} The luminosity function of red sequence galaxies
	(top panel) and blue sequence galaxies (bottom panel) in the \gband\
	band. Each panel shows the luminosity function of galaxies in three
	ways: the redshift $z\sim 0.1$ luminosity function of SDSS galaxies
	(dotted histogram); the redshift $z\sim 0.1$ luminosity function of
	SDSS galaxies as it would be measured by a DEEP2-like survey (thin
	solid histogram); and the redshift $z\sim 1$ luminosity function of
	DEEP2 galaxies. Where the two SDSS luminosity functions agree
	indicates where the DEEP2 luminosity function is unaffected by the
	explicit selection effects in DEEP2. }
\end{figure}

\begin{figure}
\figurenum{\fignum}
\plotone{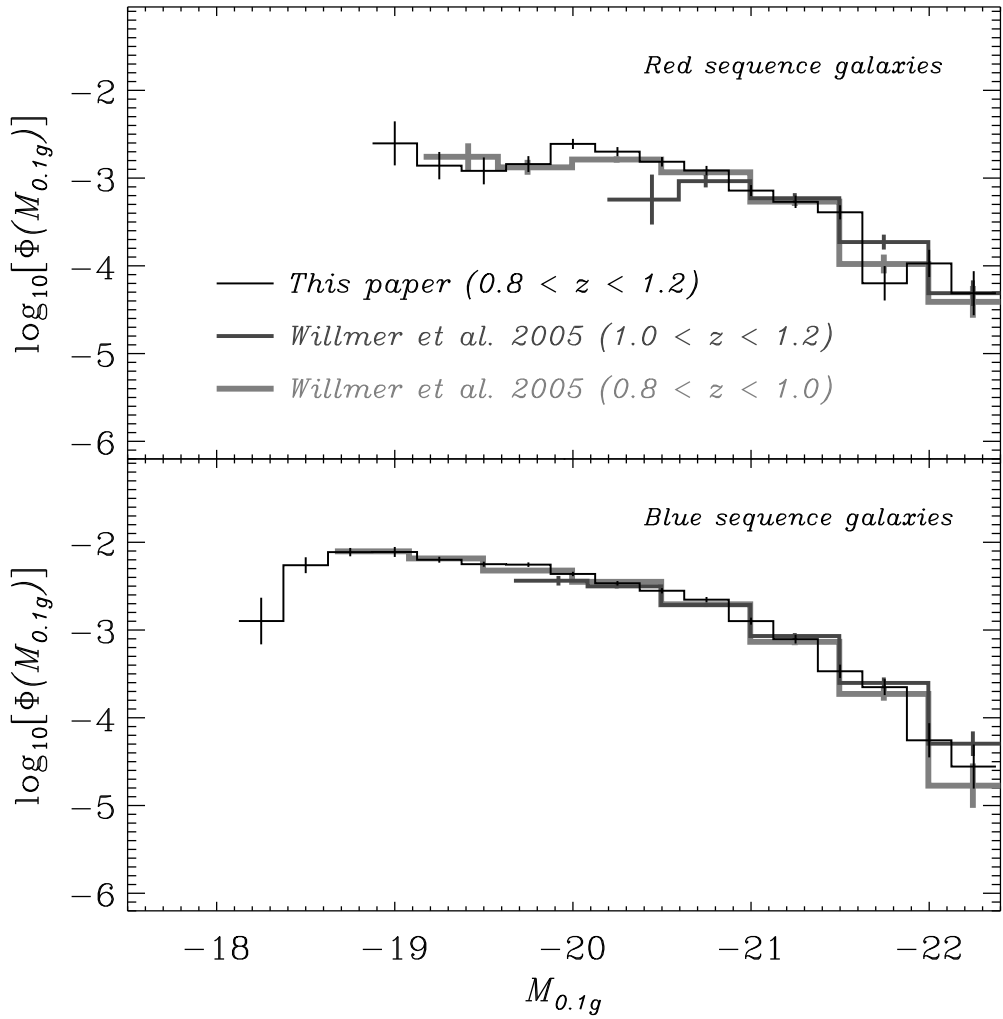}
\caption{\label{lfs_compare} Similar to Figure \ref{lfs}, but only
	showing results for the DEEP2 survey. Our results are overlaid on
	those of \citet{willmer05a}, showing that they agree well.}
\end{figure}

Figure \ref{lfs} shows the results of this calculation, with the red
galaxy luminosity function in the top panel and the blue galaxy
luminosity function in the bottom panel. Table \ref{deep2sdss_lfs}
lists the original SDSS luminosity function and the DEEP2 luminosity
function in these figures. The original SDSS sample (dotted histogram)
and the converted SDSS sample (thin solid histogram) agree for $\absmg
 < -19$. This agreement demonstrates that in that regime
one can correct the DEEP2 survey for its explicit color cuts (whether
I have correctly described the incompletenesses that I have also
corrected for cannot be addressed by this comparison). As a further
check, Figure \ref{lfs_compare} compares these results to those of
\citet{willmer05a} (and also in \citealt{faber05a}, and good agreement
is found. Similarly good agreement is found between both these results
and the COMBO-17 results of \citet{bell04a}, as \citet{faber05a} show.

\begin{figure}
\figurenum{\fignum}
\plotone{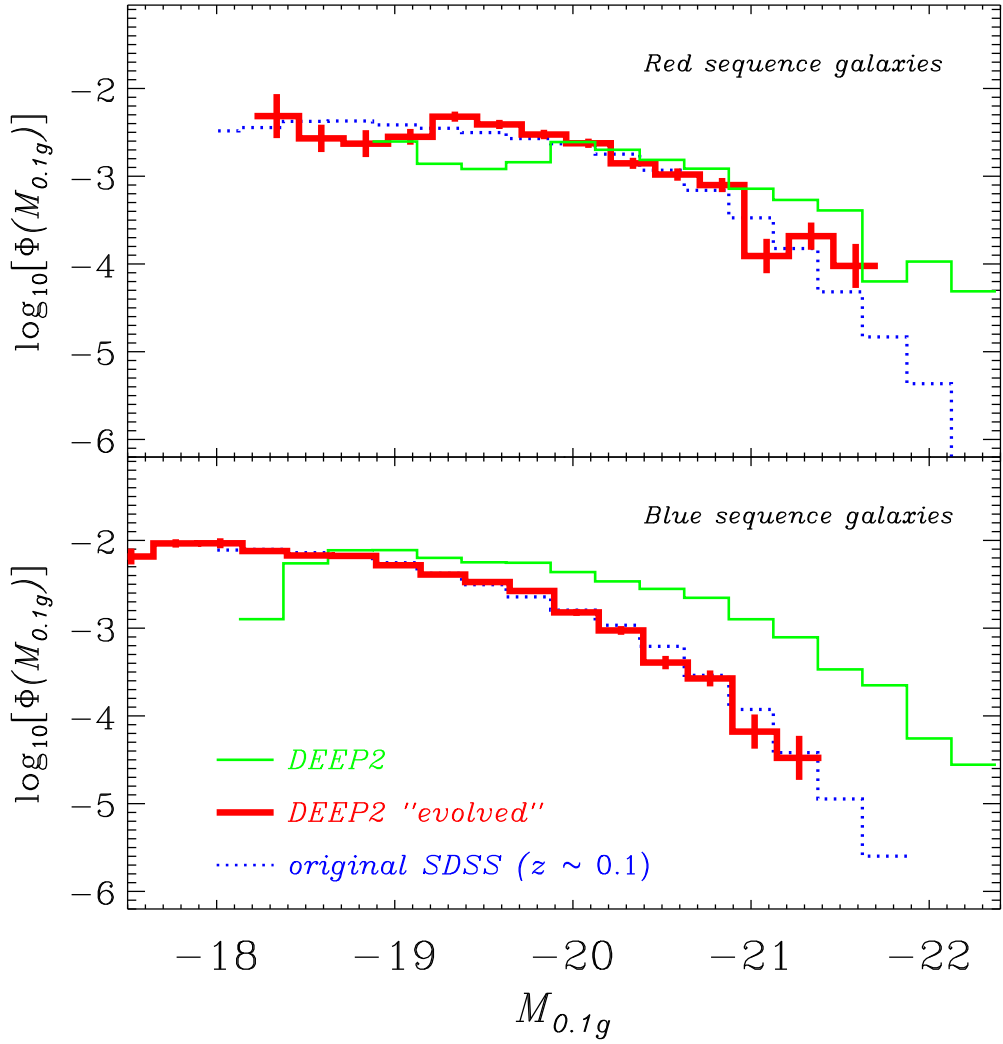}
\caption{\label{lfs_shifts} Similar to Figure \ref{lfs}. In addition
	to the original DEEP2 luminosity function (thin solid histogram), I
	show that luminosity function shifted according the parameters
	listed in Table \ref{lf_shifts} (thick solid histogram). The result
	matches the SDSS luminosity function reasonably well.}
\end{figure}

One can explore the differences between these luminosity functions in
two ways. First, one can simply ask how much one needs to shift the
absolute magnitudes and the overall number density at $z\sim 1$ to
match the luminosity functions at $z \sim 0.1$, keeping the shape of
the luminosity function fixed. I determine the optimal shift $\Delta
M_{\gband} = M_{\gband,\mathrm{SDSS}}- M_{\gband,\mathrm{DEEP2}}$ and
offset $f_\mathrm{off} = \rho_\mathrm{SDSS}/\rho_\mathrm{DEEP2}$ by
minimizing $\chi^2$ and estimate the uncertainties with a jackknife
technique. Table \ref{lf_shifts} lists the best-fit shifts and
uncertainties. Figure \ref{lfs_shifts} shows the resulting luminosity
function at redshift $z \sim 0.1$. The blue sequence is consistent
with a simple fading by about $1.0\pm 0.1$ mag, with a factor of $1.2
\pm 0.14$ offset in the overall number density.  On the other hand, it
is difficult to avoid a change in the overall number density of red
galaxies, as previous workers have found. The best fit is about a
factor of two. 

\begin{figure}
\figurenum{\fignum}
\plotone{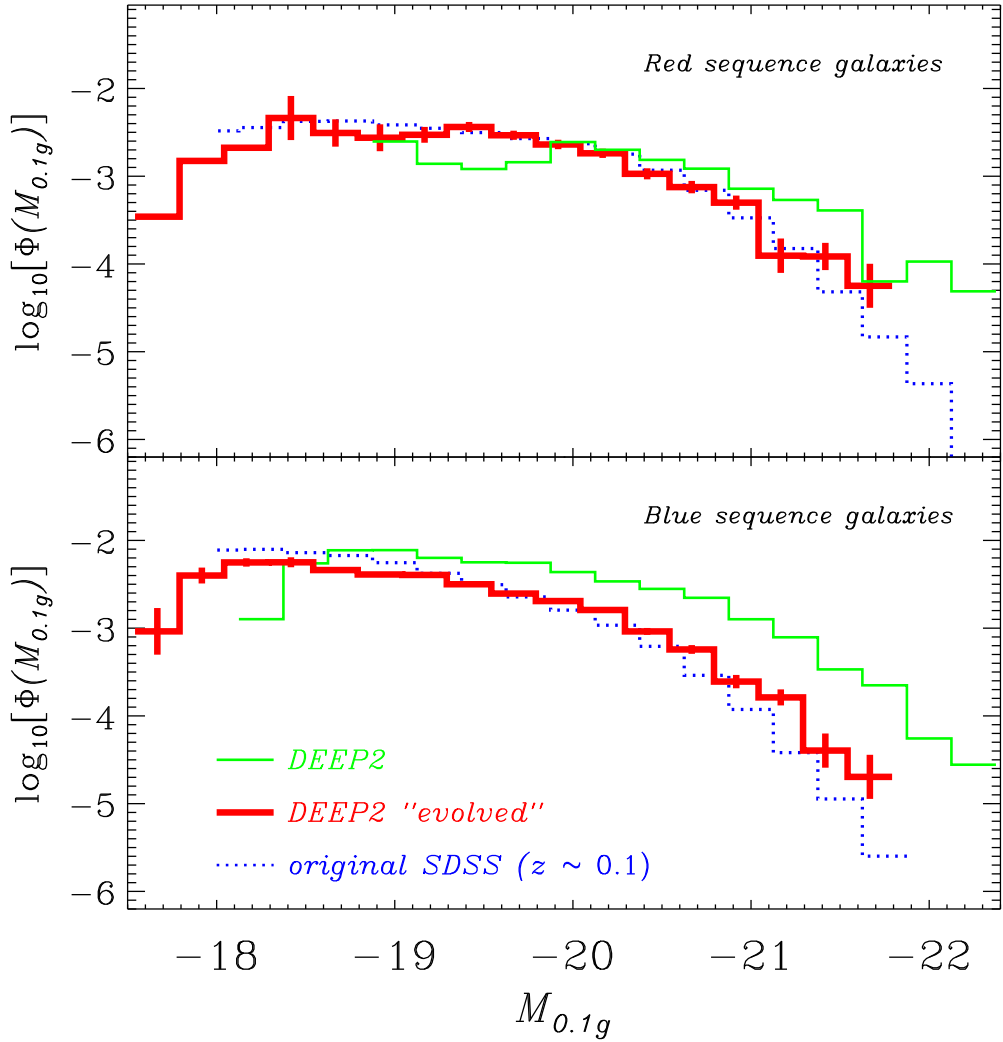}
\caption{\label{lfs_flow} Similar to Figure \ref{lfs_shifts}. Now I
	show the result of allowing both luminosity functions to fade by
	about 0.6 mag, and transferring 25\% of the blue galaxies from the
	blue to the red sequence (thick solid histogram). Again, this
	matches the SDSS luminosity function reasonably well.}
\end{figure}

Second, one can require that the numbers of galaxies be roughly
conserved by fitting for a single shift in magnitude of all galaxies,
but allowing some galaxies to transfer from the red sequence to the
blue sequence by some process. The best fit for a shift is $\Delta
M_{\gband} = 0.6 \pm 0.05$, with a fraction of blue galaxies
transferring to the red sequence of $0.25 \pm 0.05$. Figure
\ref{lfs_flow} shows the result. This picture is somewhat simpler than
the first, not requiring the majority of red galaxies at low redshift
to have formed from many mergers objects below the detection limit.
Note that the number densities of the most luminous red galaxies are
hardly affected by the transfer from the blue sequence, and that the
number densities of these objects are in agreement if the luminosity
evolution is only 0.6 mag (which is hard to get from standard passive
evolution scenarios).

\section{ A simple model of the color-magnitude diagram and the luminosity
functions}
\label{model}

In the previous section, I found that between redshift $z=1$ and
$z=0$ the galaxies on both the blue sequence and red sequence become
redder and less luminous. Here I try to interpret these results in
some simple ways.  My approach is {\it not} to hypothesize about the
fundamental physical processes at work --- which might include
accretion, ram pressure stripping, and merging, among others --- but
to simply ask what these observations may tell us about the
star-formation histories.

\subsection{Description of the model}

As a tool, I define a simple set of models for the star-formation
history of galaxies.  All of the predictions are based on spectra
produced for these models in the stellar population synthesis package
of \citet{bruzual03a} using the \citet{chabrier03a} stellar initial
mass function (IMF).

The model for the blue sequence consists of an initial burst of
duration 1 Gyr starting at the beginning of the Universe, followed by
quiescent star-formation until $z=1$ (5.8 Gyrs). I use four such
models, with initial burst star-formation rates 2, 15, 50, and 120
times the quiescent rate (to land on the blue sequence at $z\sim 1$),
and with metallicities 30\%, 50\%, 70\% and 90\% solar (to land on the
red sequence at $z\sim 0.1$ if star-formation ends).  I set the
normalizations set such that the model galaxies follow the blue
sequence at $z=1$.  After $z=1$ I allow three alternatives: a cutoff
in the star-formation, a reduction in the quiescent star-formation
rate by a factor of three, and a continuation of the quiescent
star-formation at a constant rate.

\begin{figure}
\figurenum{\fignum}
\plotone{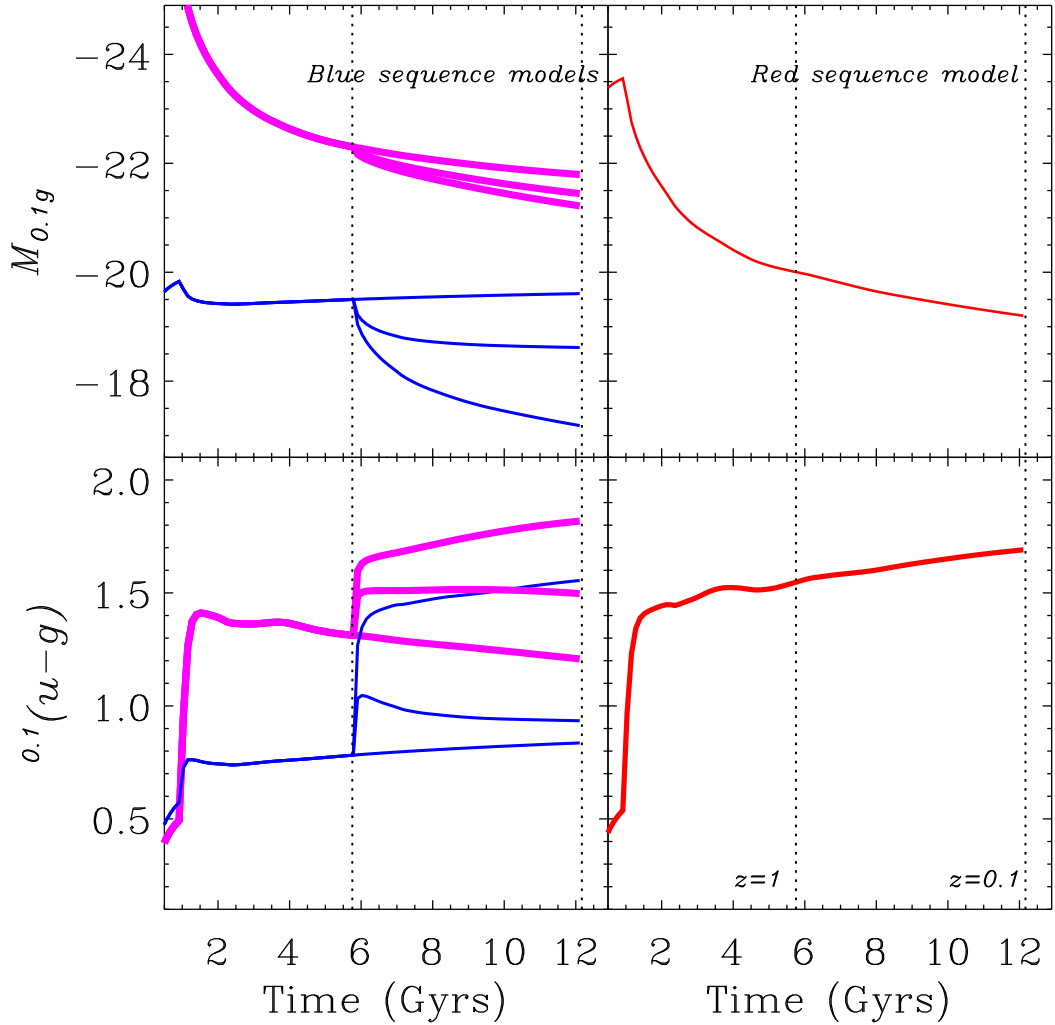}
\caption{\label{sfh_models} Colors and absolute magnitudes as a
	function of time, for models of the blue sequence (left panel) and
	red sequence (right panel) galaxies. The thick lines for the blue
	sequence galaxies correspond to a model with a large initial burst,
	the thin to a model with a small initial burst. The trifurcation at
	redshift $z=1$ indicates (from bottom to top) a continuation of
	star-formation at a constant rate, a decline in star-formation by a
	factor of three, and a complete cutoff in star-formaton.}
\end{figure}

The left-hand panels Figure \ref{sfh_models} shows the evolution in
absolute magnitude (top panel) and color (bottom panel) for blue
sequence models with burst star-formation rates of 2 (thin line) and
120 (thick line) times the quiescent rate. Each of the three
alternatives are shown for the behavior after redshift $z=1$.  In each
case, a reduction in star-formation increases the rate of fading and
reddening. A complete cutoff in star-formation puts a galaxy onto the
red sequence very rapidly, in around 1 Gyr, and leads to the most
rapid fading.  A continuation of star-formation at a constant rate
leads to nearly constant colors and absolute magnitudes. Meanwhile, a
simple reduction in the star-formation rate leads to a rapid reddening
of the color (but not all the way onto the red sequence) and a
reduction in the luminosity. For the small initial burst, the change
in absolute magnitude after a reduction of the star-formation rate is
about 0.90 mag, and for the large initial burst it is about 0.85 mag.
The changes in \umg\ color are 0.15 mag and 0.2 mag
respectively. Taken together and accounting for the slope of the color
magnitude diagram in Table \ref{cmag_table}, this makes the shift in
color at fixed absolute magnitude about 0.3, approximately that
observed.

The rapidity of the reaction to any change in star-formation rates
indicates that bimodality in the color distribution is a generic
feature arising from stellar population synthesis. As long some
galaxies can have star-formation rapidly cut off, it is difficult to
avoid a strong separation in color-space from those who have had such
a cut off and those who have not.

The right-hand panel shows my model for a typical red sequence galaxy,
a stellar population of solar metallicity born in a 1 Gyr burst at the
beginning of the Universe, passively evolving thereafter. We will see
from the examination of the blue sequence models that blue sequence
galaxies whose star-formation ends abruptly can rapidly move onto the
red sequence, so by no means am I suggesting that this red sequence
model describes the star-formation of all red sequence
galaxies. However, as long as they sit on the red sequence, it does
not matter much how they got there: their evolution in color and
magnitude is about the same. Between redshifts $z=1$ and $z=0$ the
absolute magnitude shifts by 0.80 mag and the color by about
0.15. Because the burst in this model is as early as possible, it
represents about the minimum evolution one could expect from these
models.

\subsection{Comparison to the color-magnitude diagram}

\begin{figure}
\figurenum{\fignum}
\plotone{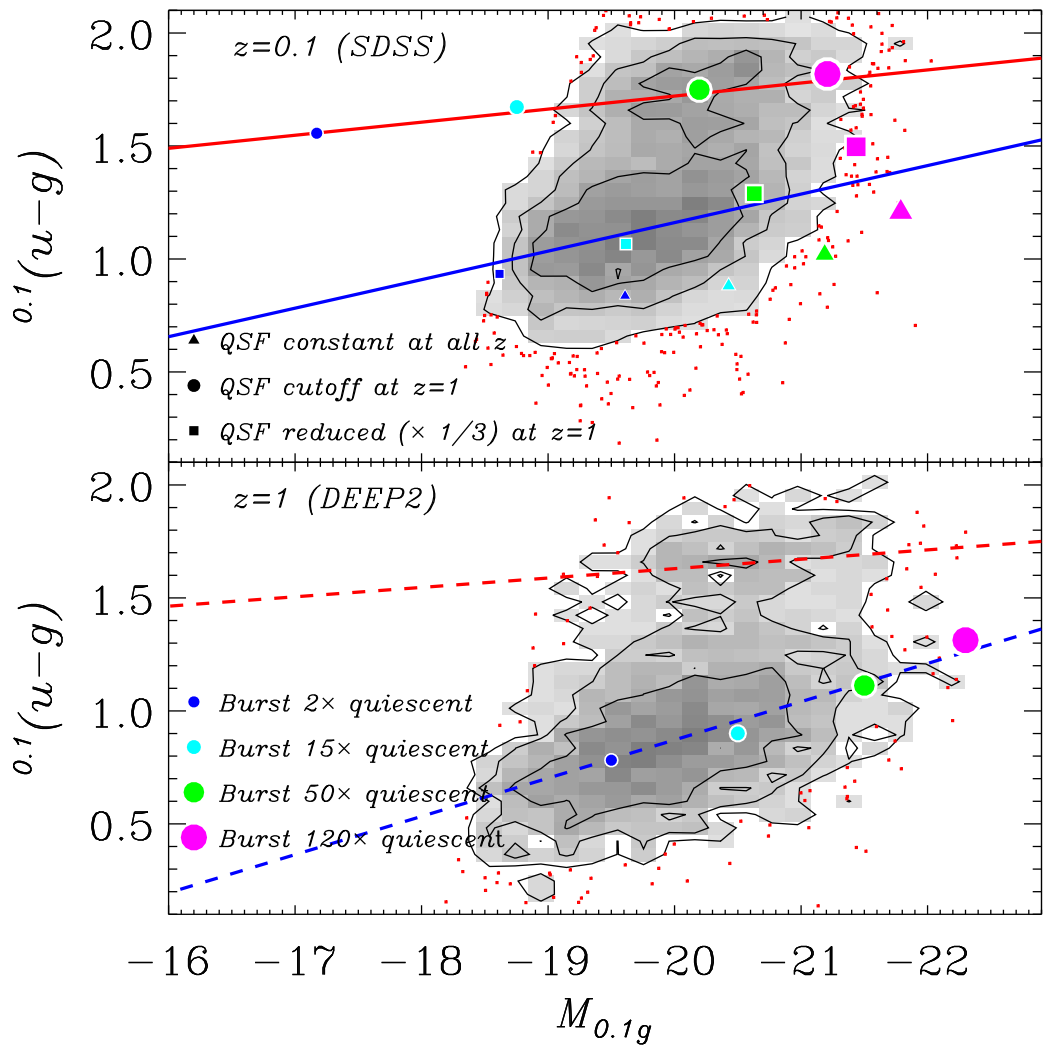}
\caption{\label{cmag_models} Same data as Figure \ref{cmag}, with some
	simple models overplotted (see Figure \ref{sfh_models}). Twelve
	models are presented in total, each consisting of a 1 Gyr burst
	starting at $t=0$ followed by quiescent star-formation (QSF) until
	at least redshift $z=1$ ($t=5.7$ Gyrs). There are four different
	burst strengths, shown with increasingly large symbols; for each
	burst strength the normalization is set to lie on the blue sequence
	at redshift $z=1$.  There are three different behaviors after
	redshift $z=1$. For the circles, the quiescent star-formation cuts
	off at $z=1$. For the squares, it is reduced by a factor of
	three. For the triangles, it continues until redshift $z=0$. Each
	panel shows the color and magnitude of each model at the given
	redshift. }
\end{figure}

Figure \ref{cmag_models} shows the results of these star-formation
histories at various points during the Universe.  The bottom panel
shows all the models at redshift $z=1$. By design, the normalizations,
initial bursts strengths, and metallicities of the blue sequence
models are set so all of the models actually do sit on the blue
sequence.

The top panel of Figure \ref{cmag_models} shows all the models at
redshift $z=0$. For the blue sequence models, the figure shows the
results of the different hypotheses. Obviously, the case in which the
star-formation is abruptly cut off results in all of the models
sitting near the red sequence. The case in which star-formation
continues at a constant rate results in bluer colors than is
typical. The case in which star-formation declines by a factor of
three results in models sitting on the blue sequence at redshift
$z=0.1$. Obviously, I chose a factor of three to guarantee this, and
the exact level of decline depends on the metallicities and dust
contents, but it is fairly clear that the star-formation rates of
typical blue sequence galaxies has declined with time.

By design, I have chosen the metallicity of the red sequence model to
land near the red sequence at both redshift $z=1$ and $z=0$. I note
in passing, but do not dwell on, the fact that to achieve the redder
colors at the tip of the red sequence, I must assume higher
metallicities, which will evolve more quickly in color at low redshift
than observed. I choose not to dwell on this difficulty because the
detailed evolution of the colors of red galaxies almost surely depend
on their abundance of alpha-elements, in a way that nobody can yet
reliably predict.
 
\subsection{Comparison to the luminosity function}

What do these models say about the luminosity functions? From the
models, and based on the change in color, I expect the blue sequence
to fade by about 0.9 mag, and this fading is extremely close to what
is observed (if the number density is roughly fixed). For example, the
fit in Figure \ref{lfs_shifts} shows the result of fading the blue
sequence luminosity function by around this much, with about a 20\%
increase in the number density. There is therefore probably no large
fraction of galaxies lost from the blue sequence due to catastrophic
events such as mergers or ram pressure stripping. For example, if I
impose a fixed shift of 0.9 mag on the blue sequence, and fit for the
scale change in the number density ($f_\mathrm{off}$) I find $1.1\pm
0.1$, as listed in Table \ref{lf_shifts}.

On the red sequence, the passive fading of almost any population,
regardless of metallicity or previous star-formation history, is about
0.80 mag. Table \ref{lf_shifts} shows that such fading produces a red
luminosity function with about 50\% the normalization of the low
redshift red galaxy luminosity function. As \citet{bell04a} conclude,
if one thinks that red galaxies fade passively, their number densities
must be boosted considerably over time by some process.

In general, these models predict more passive evolution than the red
sequence luminosity function evolution can tolerate. In addition,
because these models invoke a very old burst, they represent a minimum
passive evolution rate (given the IMF choice).  Thus, if one trusts
these models and the relative absolute photometry at the 0.2 mag
level, these results require growth along the red sequence due to
mergers, low level star-formation, or migration from the blue
sequence.

\subsection{Caveats to the comparisons}

Of course, the model here is extremely simplistic but it would not pay
to create a more baroque model given the data set I use here. For
example, the metallicity of galaxies is definitely a function of
absolute magnitude, but this change has little effect on the
prediction of the \umg\ colors of the star-forming galaxies on the
blue sequence. Another way of stating this fact is that if I imposed a
mass-metallicity relationship in the models, I would simply have
chosen different burst strengths to put the model galaxies on the blue
sequence at redshift $z=1$.  The metallicity does affect where the
galaxies end up on the red sequence, but having only the \umg\ color
gives us no handle on what these metallicities really are anyway.

In addition, the change in location of the blue sequence may not be a
change in star-formation rate, but simply a difference in the amount
of dust in the galaxies.  For the Milky Way dust law, 1 mag of
extinction in the \gband-band reddens galaxies by 0.3 mag in
\umg. Given that star-forming galaxies today are believed to have
internal extinction of order 1 mag, if {\it all} of the change in
color of the sequence is due to an increase in the dust content it
implies that galaxies at redshift $z=1$ have near zero dust extinction
--- which is possible but unlikely.

There are observational caveats, as well.  For a number of reasons,
the DEEP2 and SDSS magnitudes are not necessarily comparable. First,
this comparison requires the AB calibration for both surveys to be good
to significantly better than 0.05 mag, which I consider here to be
likely.  Second, the galaxy photometry is also not necessarily
consistent between the surveys. Although the colors are likely to be
robust, at the 0.05 mag level the overall normalization of galaxy
magnitudes can depend on the method of calculation. This problem is
probably most acute on the red sequence, where there are a significant
number of de Vaucouleurs galaxies (for exponential galaxies most
measures of flux contain all of it).  The DEEP2 survey used aperture
magnitudes with radii that for redshift $z=1$ galaxies correspond to
about $3r_{50}$ ($r_{50}$ is typically 3--4 $h^{-1}$ kpc for
reasonably luminous red galaxies, and the angular size of the
apertures were about 1$''$.5), which is close to the same as the SDSS
Petrosian aperture.  The difference, empirically, between the SDSS
Petrosian fluxes and the SDSS model fluxes (used here) is about 5\%,
in the sense that the model fluxes are slightly brighter. Therefore, I
am reasonably confident that the DEEP2 and SDSS magnitudes are
consistent at the required level.

A second observational caveat is that there may be selection effects
eliminating low luminosity red galaxies preferentially, since they
tend to be low in surface brightness. Since I do account for the
spectroscopic incompleteness as a function of color and magnitude, the
most likely culprit here would be a dependence of the photometric
completeness on surface brightness.  Such a dependence could
preferentially bias us against the lowest luminosity red galaxies the
most. Short of reimaging the area in question or reanalyzing the
photometry itself, one does not have much hope of estimating the
magnitude of such an effect or determining if it exists.

Finally, a large uncertainty is simple cosmic variance.  The high
redshift sample only consists of a volume of about a few times $10^5$
$h^{-3}$ Mpc$^3$ (and smaller than that for the fainter galaxies in
this sample). Such a volume only one or two large clusters in it even
at low redshift. Thus, I don't expect that the sample will span the
range of cosmic structures or even necessarily constitute a fair
sample yet.

\section{ Discussion}
\label{discussion}

\subsection{ A quiet life for the blue sequence}

The most striking result here is the mild evolution of the blue
sequence, which is consistent with a constant number density to within
$\sim 10\%$, and a simple decline of the star-formation rate by an
amount which is consistent with the change in the color of the blue
sequence.

The mild evolution of the blue sequence appears inconsistent with both
theoretical predictions and observations of merger rates that suggest
that $> 50\%$ of all galaxies undergo a merger between redshifts $z=1$
and $z=0$ (\citealt{baugh96a, lefevre00a, bell04a, bell05a}).  If such
mergers cause galaxies to leave the blue sequence, between redshifts
$z=1$ and $z=0.1$ the blue sequence would decline a factor of two,
which my results suggest it does not.  To reconcile the results here
with such a high merger rate requires replenishing the blue sequence
from some other source (such as former {\it red} sequence galaxies
that acquire new gas?).

On the other hand, other direct observational measurements of the
merger histories suggest relatively low numbers of recent mergers,
closer to 10\% (\citealt{patton02a, bundy04a, lin04a}). I will not
attempt here to reconcile various measurements of the numbers of
observed mergers in the Universe.  Suffice it to say that measuring
pair counts and their change over time is fraught with systematic
worries, such as projection effects, Malmquist-type biases, and
$K$-correction related selection effects.  Furthermore, estimating
merger rates from pair counts involves assumptions about the time
scales of the mergers.

My results here are much more indirect than direct counting of pairs,
but suggest that if mergers are catastrophic events for blue galaxies,
there cannot have been a large number of them, favoring those results
which find a low merger rate (for blue galaxies).

\subsection{ Merging probably required on the red sequence}

Meanwhile, the red sequence appears to grow considerably in this time
period. Two methods of growing the red sequence seem possible: the
migration of blue galaxies to the red sequence due to a cut-off in
their star-formation and dry mergers between red galaxies.

As Figure \ref{sfh_models} shows, a sharp cutoff in star-formation of
a blue sequence galaxies allows it to migrate within 1 Gyr to the red
sequence.  To get enough galaxies to undergo this process and populate
the faint end of the red sequence at redshift $z=0.1$ one needs 25\%
of the blue population to migrate, which is very marginally
allowed. However, to get enough galaxies to populate the bright end of
the red sequence at redshift $z=0.1$ one cannot rely on migration from
the blue sequence, since there are not enough luminous blue galaxies
at redshift $z=1$. Instead, one of three things must be true: passive
evolution is 0.2 mag slower than the scenarios I present here, which
are about as slow as current models allow; the fluxes of the SDSS and
DEEP2 galaxies are inconsistent in an absolute sense by 0.2 mag, which
is unlikely; or the luminous red galaxies grow by dry mergers.

In the case of the red galaxies, high merger rates in the observations
improve the consistency checks. For example, \citet{bell05a} estimate
based on seven observed close pairs between $0 <z <0.7$ that every
galaxy on the red sequence merges about once, which may be enough to
do the job.

\section{ Conclusions}
\label{conclusions}

I have presented a comparison of the low redshift ($z\sim 0.1$) SDSS
data with the high redshift ($z\sim 1$) DEEP2 data. In general, the
data suggest a quiet life for most galaxies on the blue sequence, with
fewer than 10\% being destroyed between those two epochs. It remains
for better quantitative predictions of $\Lambda$CDM than currently
exist to determine whether that theory is consistent with this
relatively peaceful history.

In particular, the data alone indicate that:
\begin{enumerate}
\item the red sequence of galaxies is redder by about 0.1 mag in the
	low redshift data; 
\item the blue sequence of galaxies is redder by about 0.3 mag in
	the low redshift data; 
\item the change in the blue sequence luminosity function is
	consistent with pure fading; and
\item the change in the red sequence luminosity function requires
	something more complex, perhaps an increase in number density.
\end{enumerate}

I have further interpreted these results in terms of simple models
for the star-formation histories of galaxies. From these models I
conclude:
\begin{enumerate}
\item that galaxies migrate quickly (1 Gyr) to the the red sequence
	when their star formation is sharply cut off; 
\item the colors and luminosities of blue sequence galaxies are
consistent with a decline in star-formation rate by about a factor of
three in the last 8 Gyrs; 
\item that consistency on the blue sequence suggests that to within
	$\sim 10\%$ mergers do not reduce the blue sequence of galaxies,
	contradicting predictions but in less obvious contradiction to
	direct observations of mergers; and
\item absent errors in the models or photometry at the 0.2 mag level,
	the data requires mergers to produce the luminous end of the red
	sequence.
\end{enumerate}

It is worth contrasting the change of the galaxy population with time
to the change of the population with environment, since it illuminates
a mystery regarding the latter. As \citet{hogg04a} and others find,
the red and blue sequences of galaxies do not change with environment,
they are simply differently populated. In contrast, I find here that
the red and blue sequences do change with time. Thus, the analogy
often invoked that underdense regions are in some sense ``younger''
regions of the Universe is not a good analogy --- we can observe the
Universe when it was younger and it looks quite different than do
underdense regions today. This result supports the notion that the
blue galaxies in dense regions had on average about the same formation
epoch and subsequent history as blue galaxies in underdense regions.

\acknowledgments

Funding for the DEEP2 survey has been provided by NSF grant
AST-0071048 and AST-0071198.

(Some of) The data presented herein were obtained at the W.M. Keck
Observatory, which is operated as a scientific partnership among the
California Institute of Technology, the University of California and
the National Aeronautics and Space Administration. The Observatory was
made possible by the generous financial support of the W.M. Keck
Foundation. The DEEP2 team and Keck Observatory acknowledge the very
significant cultural role and reverence that the summit of Mauna Kea
has always had within the indigenous Hawaiian community and appreciate
the opportunity to conduct observations from this mountain. 

Funding for the creation and distribution of the SDSS Archive has been
provided by the Alfred P. Sloan Foundation, the Participating
Institutions, the National Aeronautics and Space Administration, the
National Science Foundation, the U.S. Department of Energy, the
Japanese Monbukagakusho, and the Max Planck Society. The SDSS Web site
is http://www.sdss.org/.

The SDSS is managed by the Astrophysical Research Consortium (ARC) for
the Participating Institutions. The Participating Institutions are The
University of Chicago, Fermilab, the Institute for Advanced Study, the
Japan Participation Group, The Johns Hopkins University, the Korean
Scientist Group, Los Alamos National Laboratory, the
Max-Planck-Institute for Astronomy (MPIA), the Max-Planck-Institute
for Astrophysics (MPA), New Mexico State University, University of
Pittsburgh, University of Portsmouth, Princeton University, the United
States Naval Observatory, and the University of Washington.

\newpage

\clearpage
\clearpage

\setcounter{thefigs}{0}

\clearpage
\stepcounter{thefigs}

\clearpage
\stepcounter{thefigs}

\clearpage
\stepcounter{thefigs}

\clearpage
\stepcounter{thefigs}

\clearpage
\stepcounter{thefigs}

\clearpage
\stepcounter{thefigs}

\clearpage
\stepcounter{thefigs}

\clearpage
\stepcounter{thefigs}

\clearpage
\stepcounter{thefigs}

\clearpage
\stepcounter{thefigs}

\newpage
\clearpage
\begin{deluxetable}{rrrr}
\tablewidth{0pt}
\tablecolumns{4}
\tablecaption{\label{cmag_table} Fits to color-magnitude relations} 
\tablehead{ Survey & Color range & $a_0$ & $a_1$ }
\startdata
SDSS & $\band{0.1}{(u-g)} > 1.5$  & $1.72$ & $-0.06$ \cr
---  & $\band{0.1}{(u-g)} < 1.5$ & $1.16$ & $-0.13$ \cr
DEEP2 & $\band{0.1}{(u-g)} > 1.4$ & $1.63$ & $-0.04$ \cr
---  & $\band{0.1}{(u-g)} < 1.4$ & $0.87$ & $-0.17$ \cr
\enddata
\tablecomments{Fits are of the form: $\umg = a_0 + a_1 (\absmg
+ 20)$.}
\end{deluxetable}

\begin{deluxetable}{rrrrr}
\tablewidth{0pt}
\tablecolumns{5}
\tablecaption{\label{deep2sdss_lfs} Luminosity functions} 
\tablehead{ $\absmg$ &
$\Phi$ (red, SDSS) & 
$\Phi$ (blue, SDSS) & 
$\Phi$ (red, DEEP2) & 
$\Phi$ (blue, DEEP2)}
\startdata
-23.50 & --- & --- & --- & --- \cr
-23.25 & --- & --- & --- & --- \cr
-23.00 & --- & --- & --- & --- \cr
-22.75 & $(2.12 \pm 2.12) \times 10^{-7}$ & --- & --- & --- \cr
-22.50 & $(2.14 \pm 2.14) \times 10^{-7}$ & --- & --- & --- \cr
-22.25 & $(1.70 \pm 1.70) \times 10^{-7}$ & --- & $(488.11 \pm 282.00) \times 10^{-7}$ & $(278.03 \pm 160.74) \times 10^{-7}$ \cr
-22.00 & $(4.32 \pm 0.82) \times 10^{-6}$ & --- & $(106.62 \pm 38.12) \times 10^{-6}$ & $(55.38 \pm 24.84) \times 10^{-6}$ \cr
-21.75 & $(1.47 \pm 0.15) \times 10^{-5}$ & $(0.25 \pm 0.06) \times 10^{-5}$ & $(6.33 \pm 2.86) \times 10^{-5}$ & $(22.35 \pm 4.69) \times 10^{-5}$ \cr
-21.50 & $(4.81 \pm 0.26) \times 10^{-5}$ & $(1.13 \pm 0.12) \times 10^{-5}$ & $(40.74 \pm 7.56) \times 10^{-5}$ & $(33.86 \pm 5.92) \times 10^{-5}$ \cr
-21.25 & $(1.50 \pm 0.04) \times 10^{-4}$ & $(0.38 \pm 0.02) \times 10^{-4}$ & $(5.37 \pm 0.87) \times 10^{-4}$ & $(7.87 \pm 0.90) \times 10^{-4}$ \cr
-21.00 & $(3.36 \pm 0.07) \times 10^{-4}$ & $(1.18 \pm 0.04) \times 10^{-4}$ & $(7.21 \pm 1.05) \times 10^{-4}$ & $(12.63 \pm 1.15) \times 10^{-4}$ \cr
-20.75 & $(6.91 \pm 0.09) \times 10^{-4}$ & $(2.90 \pm 0.06) \times 10^{-4}$ & $(12.20 \pm 1.48) \times 10^{-4}$ & $(22.17 \pm 1.54) \times 10^{-4}$ \cr
-20.50 & $(1.17 \pm 0.01) \times 10^{-3}$ & $(0.62 \pm 0.01) \times 10^{-3}$ & $(1.53 \pm 0.19) \times 10^{-3}$ & $(2.80 \pm 0.19) \times 10^{-3}$ \cr
-20.25 & $(1.79 \pm 0.01) \times 10^{-3}$ & $(1.08 \pm 0.01) \times 10^{-3}$ & $(2.00 \pm 0.24) \times 10^{-3}$ & $(3.41 \pm 0.20) \times 10^{-3}$ \cr
-20.00 & $(2.35 \pm 0.02) \times 10^{-3}$ & $(1.60 \pm 0.02) \times 10^{-3}$ & $(2.45 \pm 0.32) \times 10^{-3}$ & $(4.35 \pm 0.25) \times 10^{-3}$ \cr
-19.75 & $(2.68 \pm 0.02) \times 10^{-3}$ & $(2.27 \pm 0.02) \times 10^{-3}$ & $(1.45 \pm 0.30) \times 10^{-3}$ & $(5.56 \pm 0.33) \times 10^{-3}$ \cr
-19.50 & $(3.15 \pm 0.03) \times 10^{-3}$ & $(3.12 \pm 0.03) \times 10^{-3}$ & $(1.21 \pm 0.43) \times 10^{-3}$ & $(5.62 \pm 0.37) \times 10^{-3}$ \cr
-19.25 & $(3.52 \pm 0.03) \times 10^{-3}$ & $(4.21 \pm 0.04) \times 10^{-3}$ & $(1.39 \pm 0.50) \times 10^{-3}$ & $(6.32 \pm 0.48) \times 10^{-3}$ \cr
-19.00 & $(3.84 \pm 0.04) \times 10^{-3}$ & $(5.57 \pm 0.06) \times 10^{-3}$ & $(2.49 \pm 1.44) \times 10^{-3}$ & $(7.75 \pm 1.01) \times 10^{-3}$ \cr
-18.75 & $(4.26 \pm 0.05) \times 10^{-3}$ & $(6.72 \pm 0.09) \times 10^{-3}$ & --- & $(7.72 \pm 0.85) \times 10^{-3}$ \cr
-18.50 & $(4.22 \pm 0.06) \times 10^{-3}$ & $(7.25 \pm 0.12) \times 10^{-3}$ & --- & $(5.47 \pm 1.14) \times 10^{-3}$ \cr
-18.25 & $(3.59 \pm 0.07) \times 10^{-3}$ & $(7.91 \pm 0.18) \times 10^{-3}$ & --- & $(1.27 \pm 0.77) \times 10^{-3}$ \cr
-18.00 & $(3.29 \pm 0.09) \times 10^{-3}$ & $(7.76 \pm 0.29) \times 10^{-3}$ & --- & --- \cr
\enddata
\tablecomments{ $\Phi$ is number per $h^{-3}$ Mpc$^3$ per unit
	magnitude. In SDSS, the division between red and blue is at $\umg =
	1.5$ and in DEEP2 it is at $\umg = 1.4$.}
\end{deluxetable}

\begin{deluxetable}{cccc}
\tablewidth{0pt}
\tablecolumns{4}
\tablecaption{\label{lf_shifts} Shifts in luminosity functions}
\tablehead{ Colors & $\Delta M_{\gband} = M_{\gband,\mathrm{SDSS}}-M_{\gband,\mathrm{DEEP2}}$ & 
$f_\mathrm{off} = \rho_\mathrm{SDSS}/\rho_\mathrm{DEEP2}$ & $r$ }
\startdata
Blue & $0.98 \pm 0.09$ & $1.20 \pm 0.14$ & $0.45$ \cr
Blue (shift fixed) & $0.90$ & $1.13 \pm 0.13$ & --- \cr
Red & $0.66 \pm 0.04$ & $1.95 \pm 0.20$ & $0.28$ \cr
Red (shift fixed) & $0.80$ & $2.14 \pm 0.22$ & --- \cr
\enddata
\tablecomments{The ``shift fixed'' cases have $\Delta M_{\gband}$ fixed to a
	reasonable value based on stellar population synthesis models. $r$
	is the correlation coefficient between the errors in the shift
	$\Delta M_{\gband}$ and the offset $f_\mathrm{off}$.}
\end{deluxetable}

\end{document}